\documentclass[12pt]{iopart}
\usepackage{iopams}
\usepackage{lscape}
\usepackage{geometry}
\usepackage{graphicx} 
\usepackage{longtable}
\usepackage{footnote}

\begin{document}

\title[]{Measurements and
identifications of extreme ultraviolet spectra of highly-charged Sm
and Er}

\author{Y.A. Podpaly, J.D. Gillaspy, J. Reader, Yu. Ralchenko}

\address{National Institute of Standards and Technology, Gaithersburg, MD
20899, USA}

\ead{yuri.podpaly@nist.gov}

\begin{abstract}

We report spectroscopic measurements of highly charged samarium and erbium
performed at the National Institute of Standards and Technology (NIST)
Electron Beam Ion Trap (EBIT). These measurements are in the extreme
ultraviolet (EUV) range, and span electron beam energies from  0.98 keV to
3.00 keV. We observed 71 lines from Kr-like Sm$^{26+}$ to Ni-like Sm$^{34+}$,
connecting 83 energy levels, and 64 lines from Rb-like Er$^{32+}$ to Ni-like
Er$^{40+}$, connecting 78 energy levels. Of these lines, 64 in Sm and 60 in Er
are new. Line identifications are performed using collisional-radiative
modeling of the EBIT plasma. All spectral lines are assigned individual
uncertainties, most in the $\sim$0.001 nm range. Energy levels are derived
from the wavelength measurements. 

\end{abstract}


\maketitle

\section{Introduction}
\label{sec:introduction}

Spectroscopy of rare earth elements has recently become a subject of active
research due to the possible use of gadolinium and terbium as next generation
light sources for extreme ultraviolet (EUV) lithography
\cite{PhysRevA.86.042503, 1402-4896-80-4-045303}. There are also few available
data about transitions of highly charged ions in the lanthanides, which makes
this an important area for additional study. 

In the NIST Atomic Spectra Database \cite{NIST_ASD}, erbium and samarium
transitions are available primarily for Sm I and II and Er I, II, and III.
Fewer data are available for more highly ionized ions. Some transitions have
been measured on Electron Beam Ion Traps (EBITs) in the EUV and X-ray regimes
\cite{PhysRevA.87.062503,Beiersdorfer_AIP1993,1402-4896-2001-T92-026,
PhysRevA.63.042513}. Much of the available data have been generated from laser
produced plasmas \cite{Louzon:09, Daido:99,1402-4896-24-4-008,
PhysRevLett.45.609, Acquista:84,Ekberg:87, Doschek:88,
Ekberg:88,1402-4896-1999-T83-005}. Highly charged samarium has also been
observed in tokamak plasmas \cite{Sugar:93_AG, Sugar:93_PD, PhysRevA.38.288}.
In highly charged erbium, likewise, there are limited available data,
including that generated by EBITs \cite{PhysRevA.87.062503,
PhysRevA.51.1683,PhysRevA.52.2689}, laser plasmas
\cite{1402-4896-24-4-008,PhysRevLett.45.609,Acquista:84, Reader:83}, and
tokamaks \cite{Sugar:93_AG, Sugar:93_PD,Sugar:93_Rh}. 

In this paper, we report Er and Sm n=4-n=4 transitions in the EUV, continuing
on our previous studies of Gd \cite{PhysRevA.86.042503} and Dy
\cite{kilbane_dysprosium}. A full list of identifications, wavelengths, and
wavelength uncertainties is generated, and we calculate energy levels with
uncertainties for these ions as well. The intent of this research is to expand
the number of measured transitions among the rare earth elements near those of
interest for EUV light sources and provide a systematic accounting of
uncertainties for transitions and energy levels. 

\section{Experiment}
\label{sec:experiment}

This work was performed at the NIST EBIT \cite{nist_ebit_first_res}. Sm
spectra were studied at twelve electron beam energies between 0.98 keV and 2.2
keV, and Er spectra were studied at twelve beam energies between 1.3 keV and
3.0 keV. These energies are sufficient to produce ions between approximately
Rb-like and Ni-like ionization stages \cite{NIST_ASD}. Beam currents varied
between 15 mA and 86 mA. Plasma confinement was achieved, as usual, through
the electrostatic trapping via the electron beam, two drift tubes (at 500 V
and 220 V), and a 2.8 T axial magnetic field. Er and Sm were injected into the
trap by using a multi-cathode Metal Vapor Vacuum Arc (MeVVA)
\cite{NIST_MeVVA}. The trap was emptied and new ions were injected from the
MeVVA every 10 seconds. Elements used for calibration were introduced by the
MeVVA, by the gas injection system (described in \cite{PRA_fahy}), or were
present as intrinsic impurities.

Spectra were recorded with a spectrometer designed for use in the EUV
\cite{blagojevic:083102}. The spectrometer is a flat-field variable-line
spacing type grating spectrometer. Data were collected with a 2048
pixel$\times$512 pixel (13.5 $\mu$m x 13.5 $\mu$m pixel dimensions)
liquid-nitrogen-cooled charge coupled device (CCD).  Spectra were taken as ten
one-minute exposures, and a cosmic ray filtering program was used to
automatically remove data that were outside of five Poisson standard
deviations of the signal, effectively removing the majority of the cosmic rays
and aberrant electronic noise. The spectral range for Sm measurements covered
approximately 4 nm to 20 nm, and for Er approximately 3 nm to 17 nm. 

Calibration of the samarium spectra was accomplished using twelve lines from
Ne$^{4+}$ through Ne$^{7+}$, one line of Fe$^{23+}$, one line of Fe$^{22+}$,
and one line of Ba$^{26+}$. Spectra of neon were taken at 2 keV and 4 keV,
iron was taken at 4 keV, and barium, which is an intrinsic impurity, was taken
at 5.8 keV. Calibration of the erbium spectra was performed using twelve lines
from Ne$^{4+}$ through Ne$^{7+}$, four lines from Xe$^{43+}$ and Xe$^{42+}$,
one line from Ba$^{45+}$, one line from O$^{4+}$, and one line from  O$^{5+}$.
Spectra of neon were taken at 2 keV and 4 keV, xenon and barium at 5.8 keV,
and oxygen at 1.8 keV.  All lines were fit with unweighted Gaussian profiles,
and uncertainties were generated for each calibration point. Third order
calibration polynomials relating wavelength to detector channel number were
calculated, and confidence intervals were generated from which calibration
uncertainties were derived. By setting the requirement that the calibration
polynomial fit had $\chi^{2}\!\approx\!n-N$, where $n$ is the number of
calibration points and $N$ is the degrees of freedom of the calibration curve
\cite{Hughes_Hase}, the systematic uncertainty was estimated. For samarium,
the systematic uncertainty was found to be 0.00055 nm; for erbium, the
systematic uncertainty was found to be 0.0010 nm. 

\begin{figure}[htp]
\centering
\includegraphics[height=0.45\textheight]{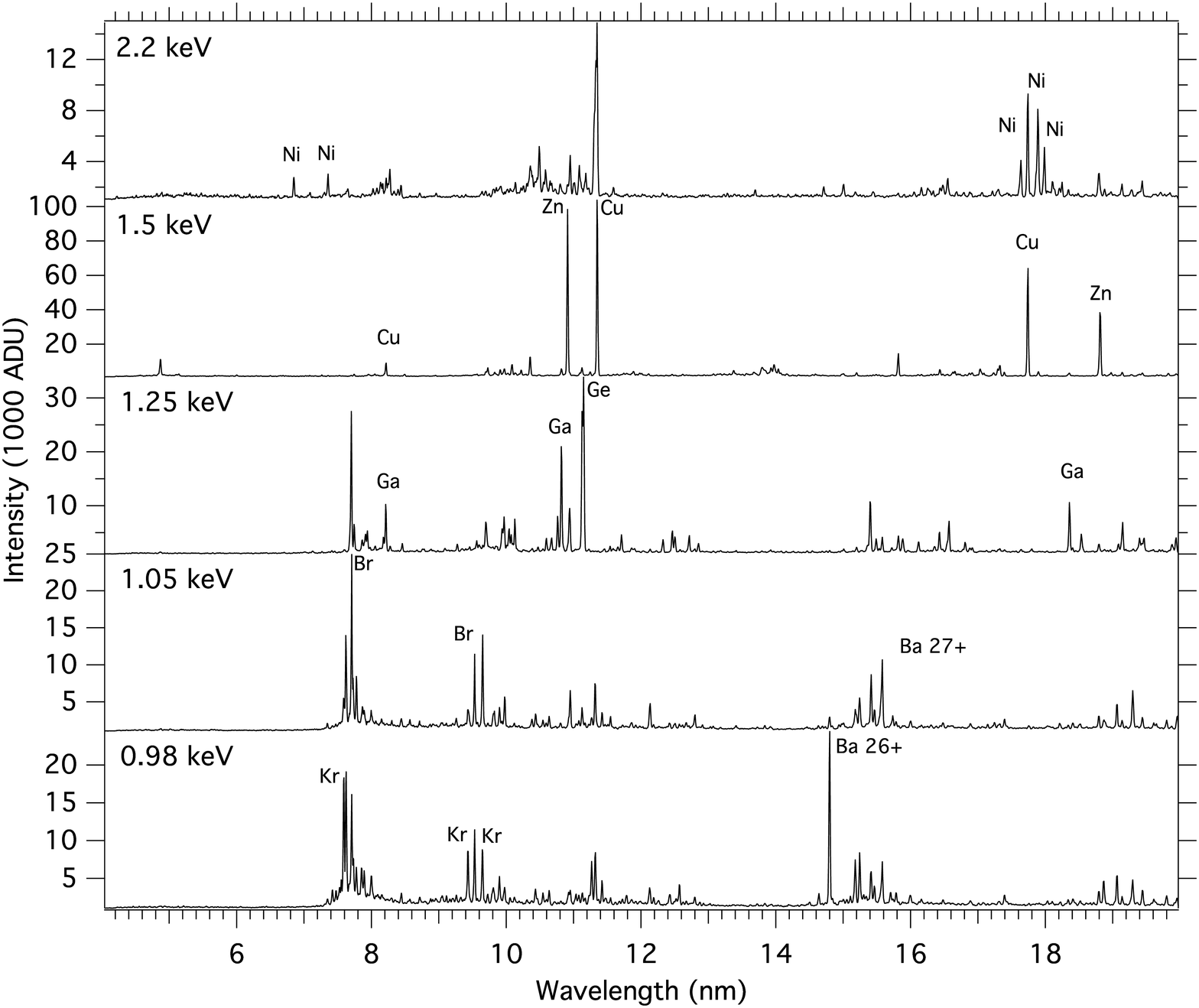}
\caption{Samarium experimental results at five beam energies. Select lines are labeled by their isoelectronic sequence.} 
\label{fig:1}
\end{figure}

\begin{figure}[htp]
\centering
\includegraphics[height=0.45\textheight]{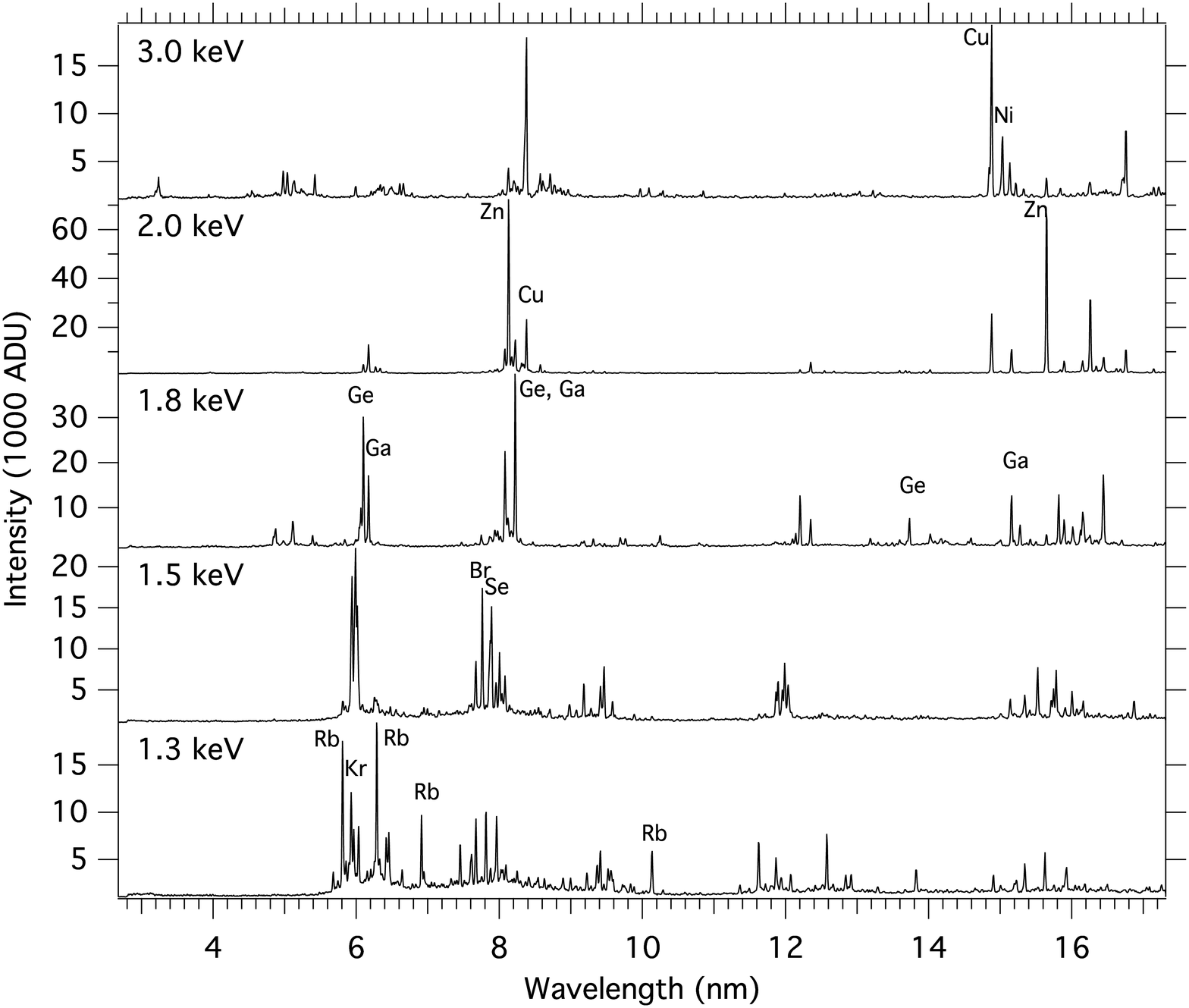}
\caption{Erbium experimental results at five beam energies. Select lines are labeled by their isoelectronic sequence.} 
\label{fig:2}
\end{figure}

Spectra were recorded at a variety of energies; typical results are shown in
figures \ref{fig:1} and
\ref{fig:2}. The intensities of the experimental spectra
are given in the analog-to-digital units (ADU) of the CCD. Lines were fit with
unweighted Gaussians, and statistical, systematic, and calibration confidence
interval uncertainties were added in quadrature for each line to generate a
total uncertainty. 

\section{Collisional-Radiative Modeling}
\label{sec:modeling}

The measured spectra were analyzed with the collisional-radiative (CR)
modeling that has been extensively described elsewhere (see, e.g.,
\cite{EUV_tungsten,DragJPB,PhysRevA.83.032517}), and therefore only the most
relevant features will be described below. The line intensities for Er and Sm
were calculated with the non-Maxwellian CR code NOMAD \cite{Ralchenko2001609}
utilizing atomic data generated with the Flexible Atomic Code (FAC)
\cite{FAC}. The level energies, radiative transition probabilities (allowed
and forbidden), and electron-impact cross sections (excitation, deexcitation,
ionization, and radiative recombination) were calculated for 8575 and 8570
levels in Sr-like to Ni-like ions of Er and Sm, respectively. The level
energies were improved using an extended calculation taking into account all
possible excitations within the n=4 complex, as described in \cite{DragJPB}.
The energies of the $3d^94l$ levels in Ni-like ions were taken from a more
accurate relativistic many-body perturbation theory (RMBPT) calculation of
\cite{Safronova200647}. The rate of charge exchange between highly-charged
ions and neutral atoms in the trap was included as the only free parameter,
aside from a small shift in electron beam energy due to space charge as
discussed below. A typical calculation of an EBIT spectrum would include 6-7
ionization stages since ions with ionization potentials larger than the beam
energy are very weakly populated. The calculated bound-bound spectra were then
Gaussian-broadened and convolved with the calculated efficiency curve of the
EUV spectrometer.

\begin{figure}[htp]
\centering
\includegraphics[height=0.5\textheight]{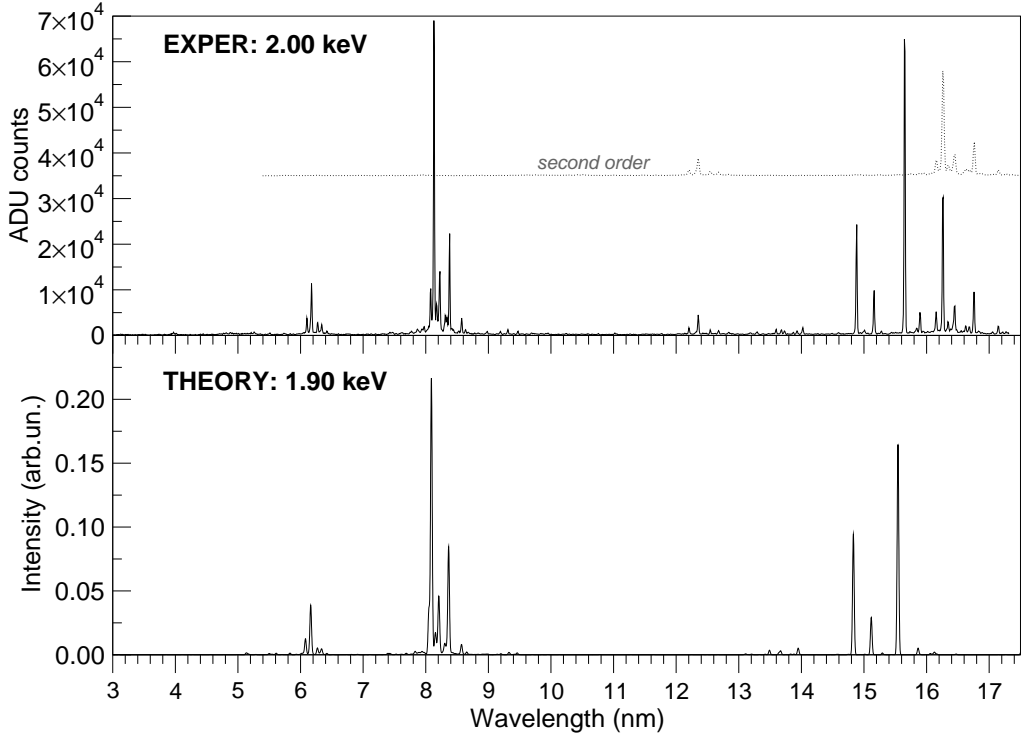}
\caption{Comparison of the experimental spectrum of erbium with the
theoretical spectrum.} 
\label{fig:3}
\end{figure}

\begin{figure}[htp]
\centering
\includegraphics[height=0.5\textheight]{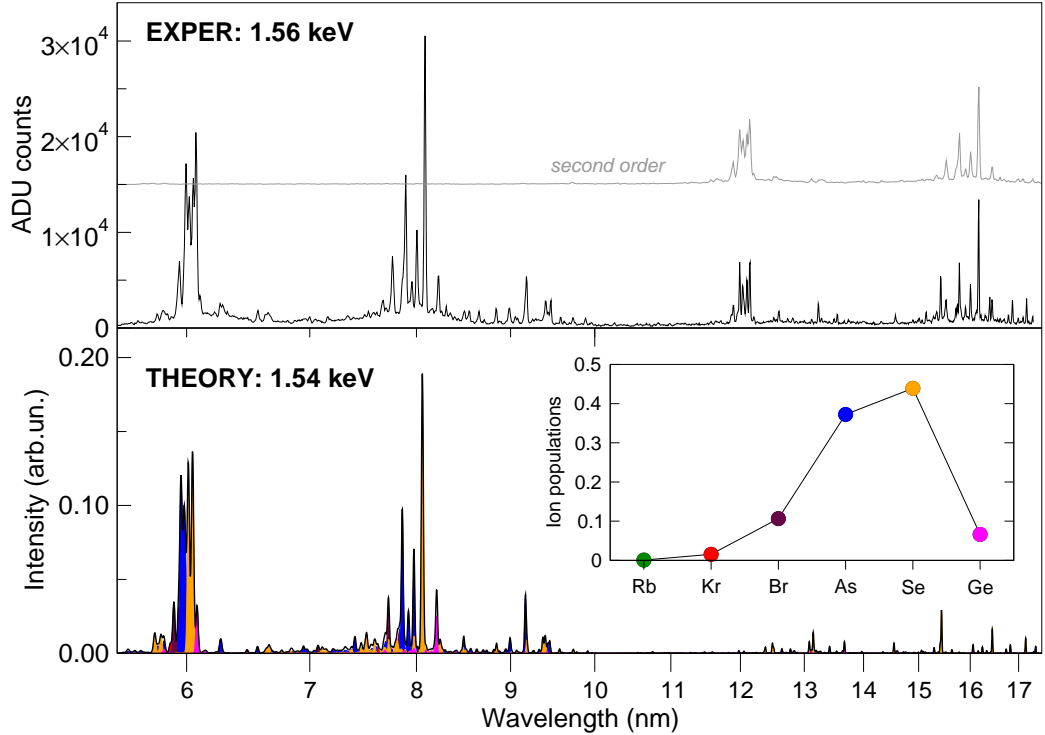}
\caption{Comparison of the experimental spectrum of erbium with the 
theoretical spectrum. The inset shows the calculated ionization distribution 
at 1.54 keV. Contributions from different ionization
stages are shown in different colors in the theoretical spectrum.} 
\label{fig:4}
\end{figure}

Examples of agreement between theoretical and experimental spectra of Er are
presented in figures \ref{fig:3} and \ref{fig:4}. The agreement between
theory and experiment for the Sm spectra is practically the same. The second
order spectra are shown by the shifted dotted lines. Starting with
figure~\ref{fig:3}, one can see that our CR modeling explains intensities
and positions of all strong spectral lines. Note that the theoretical beam
energy is lower than the nominal experimental energy; this is due to the space
charge effects that are common in EBITs. Although for some lines there is a
small shift in wavelength between theory and experiment, a very good match of
line intensities allows us to unambiguously identify all prominent lines in
the spectrum. For instance, the main groups of spectral lines near 8 nm and 15
nm are found to be due to the $4s_{1/2}$-$4p_{3/2}$ and $4s_{1/2}$-$4p_{1/2}$,
respectively, in Cu-, Zn-, and Ga-like ions of Er.

The theoretical spectrum in Fig.~\ref{fig:4} is also seen to agree well
with the measured spectrum. The contributions from different ions are shown by
different colors, while the total theoretical spectrum is marked by a black
solid line. For this energy, the calculated ion populations are the following:
[Kr]:[Br]:[As]:[Se]:[Ge] = 0.015:0.106:0.372:0.439:0.066 (see the inset in
Fig.~\ref{fig:4}), and the remaining population ($<$ 1\%) is in the lower
ions. Accordingly, the most prominent lines are due to transitions in As-like
(blue) and Ge-like (orange) ions. Although a simple visual comparison allows
us to identify practically all experimental lines, additional assistance in
identification is provided by comparison of the intensity of the lines at the
various beam energies. A similar $E_{beam}$-dependence of line intensities
from a particular ionization stage is especially helpful in identification of
blended lines.

\section{Results}
\label{sec:results}

Results of our measurements are shown in tables \ref{t:full_results_sm} and
\ref{t:full_results_er}. Wavelengths calculated by FAC are presented as well.
All uncertainties for wavelengths are reported as one standard deviation. The
table also presents previous measurements of several spectral lines. If a line has
significant blending with another line the letter `b' is appended to the
wavelength. For samarium, 71 lines were measured, with 64 of them new lines.
For erbium, 64 lines were measured, with 60 of them new lines. In the erbium
data, some the of the Br-like lines were measured from their second orders and
these wavelengths are marked with `(*2)'. 

The level identifications are given in standard notations for relativistic
configurations (e.g, $4p_+$ for $4p_{j=l+1/2}$ and $4d_-$ for $4d_{j=l-1/2}$).
As is customary for the FAC calculations, the electron pairs with total zero
momentum are omitted. Since only the largest component of a wavefunction is
presented, this may result in non-unique labels for some levels. For instance,
levels 7 and 9 in As-like Sm are both shown in table~\ref{t:full_results_sm}
as belonging to the term $4d_{-}$ of configuration $4p^24d$. The calculations
show that level~7 is composed of 45.7\% of $4p^{2}4d$ $4d_{-}$,  45.3\% of
$4s4p^{4}$ $(4s_{+}, (4p_{+}^{2})_{2})$, and 3.3\% of $4s4p^{4}$
$((4s_{+},4p_{-})_{1},(4p_{+}^{3})_{3/2})$, while for level~9 the same
relativistic terms have contributions of 49.9\%, 34.2\% and 5.3\%,
respectively. Such strong mixing is not uncommon in highly-charged N-shell
ions.

The energy levels for samarium and erbium are reported in tables
\ref{t:full_results_sm_energy} and \ref{t:full_results_er_energy}. The
conversion factor from eV to cm$^{-1}$ was 1~eV = 8065.544219(18)~cm$^{-1}$
\cite{codata}. The first energy level, ground level, is taken to have zero
energy. In some cases, groups of levels are not connected to the ground level
by measured radiative transitions, and therefore the energy of a reference
level is taken from FAC or, in the case of Ni-like ions, from
Ref.~\cite{Safronova200647}. These levels are marked in our tables with the
symbol $+$ followed by a letter. For those levels, the reported results are
given as the weighted mean ($\propto 1/\sigma^{2}$) of the possible
derivations. Since none of the derived levels have uncertainties less than 20
cm$^{-1}$, the results are all rounded to 10 cm$^{-1}$, unless they are from
calculations.

For some lines, their identifications were validated by using the Ritz
combination principle. For instance, the 8.2227~nm spectral line in Ga-like Er
that corresponds to the transition between the ground state (level 1) and
level 6, is blended by a strong line in the Ge-like ion, and therefore
additional confirmation of its wavelength would be helpful. The ground state
and the first excited level in this ion are both connected to levels 6 and 7,
all four lines having been measured in the present experiment. The energy
differences between the two lowest levels calculated from two pairs of
measured wavelengths agree within approximately 0.1\%, thereby confirming our
identifications for all those lines, including the 1--6 transition. Similar
analysis was performed for transitions in the Ni-like ions of Sm and Er.

For the spectral range analyzed here, there exist only a handful of other
measurements for Sm and Er. Our wavelengths for Ni-like Sm agree very well
with the low-resolution measurements from laser-produced plasmas
\cite{Daido:99}. The high-resolution measurements for Cu-like Sm and Er
\cite{1402-4896-24-4-008,Doschek:88} also agree with our results within
experimental uncertainties. Similar level of agreement is observed for other
measurements in Zn-like ions \cite{PhysRevLett.45.609,Acquista:84}. This
provides additional confidence in our measured wavelengths and identifications
of spectral lines from other ions of Sm and Er for which no measured lines are
known.

For most cases, the FAC wavelengths deviate from the measured values to within
a fraction of per cent. The only exception is the transitions in Ni-like Sm
that connect level 35 with quantum numbers $3d^94d$
(($3d^3_-$)$_{3/2}$,$4d_-$)$_0$ with levels 9 and 12. For this J=0 level the
difference between theory and experiment reaches 1.6\% for calculations that
include double core excitations from n=3 into n=4 (value in table
\ref{t:full_results_sm_energy}) and 4\% for single 3--4 excitations. The RMBPT
results of Ref.~\cite{Safronova200647} do not contain J=0 levels and there is
no other recent theoretical work addressing the energy of this particular
level. We would like to point out that our measured 9--35 and 12--35
wavelengths can provide an interesting test for advanced atomic structure
theories. 

\section{Conclusions}
\label{sec:conclusions}

We report the results of EUV measurements of highly charged ions of samarium
and erbium in an EBIT. One hundred and thirty-five lines are measured with
individual uncertainties calculated for each line.  From these transitions, a
total of 161 energy levels were derived and uncertainties were assigned to
each value. Overall, the agreement between theory and experiment is excellent
with respect to the intensities of the lines and very good for the
wavelengths. Our line identifications agree very well with the already known
data for Ni-, Cu-, and Zn-like ions.

\section{Acknowledgments}
\label{sec:acknowledgements}

The authors would like to thank J. Smiga for assistance with calibration of
the erbium spectra. This work was funded in part by the Office of Fusion
Energy Sciences of the U.S. Department of Energy. Y.P. is supported by a
postdoctoral appointment with the National Institute of Standards and
Technology National Research Council Research Associateship Program. 

\section*{References}

\begin{landscape}

\small

\begin{longtable}{l l l l l l l l }
\caption{Wavelengths (nm) of highly charged samarium. 
The numbers in parentheses are the wavelength uncertainties in units of 
the last significant digit.
The numbers in square
brackets in column ``FAC" are the ordinal numbers for the 
lower and upper levels. b-blended line. 
\label{t:full_results_sm}
}\\
 Stage &  \multicolumn{2}{l}{ Lower Level} &   \multicolumn{2}{l}{ Upper Level}  &  Wavelength &  FAC &  Prev. Exp.  \\
&  Conf. &  State &  Conf.  &  State &    &    &     \\ 
\hline
 34+ [Ni] & $3d^{9}4p$ & $((3d_{-}^3)_{3/2},4p_{-})_{1}$  	& $3d^{9}4d$ &  $ ((3d_{-}^3)_{3/2},4d_{-})_{0}$ 	&  6.8494(7)   & 	6.7384   [9-35]	& 	6.85(2)$^{a}$ \\ 
 34+ [Ni] & $3d^{9}4p$ &  $((3d_{+}^5)_{5/2},4p_{+})_{1}$    & $3d^{9}4d$  &  $((3d_{-}^3)_{3/2},4d_{-})_{0}$ 	&  7.3563(8) 	& 	7.2344   [12-35] & 	7.36(2)$^{a}$ \\
 34+ [Ni] & $3d^{9}4p$ & $((3d_{+})^5_{5/2},4p_{+})_{4}$ & $3d^{9}4d$ 	& $((3d_{+}^5)_{5/2},4d_{+})_{5}$  	&  10.4908(11)   &  	10.4800 [10-23] & 		   \\
 34+ [Ni] & $3d^{9}4s$&  $((3d_{+}^{5})_{5/2},4s_{+})_{3}$ & $3d^{9}4p$ & $((3d_{+}^5)_{5/2},4p_{-})_{3}$ 	&  17.6366(17) 	& 	17.5836 [2-7]	& 		   \\
 34+ [Ni] & $3d^{9}4s$ &  $((3d_{+}^{5})_{5/2},4s_{+})_{3}$& $3d^{9}4p$ &  $((3d_{+}^5)_{5/2},4p_{-})_{2}$ 	&  17.8926(14)   & 	17.8722 [2-6]	& 		   \\
 34+ [Ni] & $3d^{9}4s$ &  $((3d_{+}^5)_{5/2},4s_{+})_{2}$ & $3d^{9}4p$&  $((3d_{+}^5)_{5/2},4p_{-})_{3}$ 	&  17.9864(16)   & 	17.9661 [3-7]	& 		   \\
 34+ [Ni] & $3d^{9}4s$&  $((3d_{-}^3)_{3/2},4s_{+})_{2}$ &  $3d^{9}4p$ &  $((3d_{-}^3)_{3/2},4p_{-})_{2}$ 	&  18.1084(20)	& 	18.0985 [5-8]	& 		  \\
 34+ [Ni] & $3d^{9}4s$ &  $((3d_{+}^5)_{5/2},4s_{+})_{2}$& $3d^{9}4p$ & $((3d_{+}^5)_{5/2},4p_{-})_{2}$ 	&  18.2491(16)  & 	18.2674 [3-6]	& 		   \\
\\
 33+ [Cu]   & $4p$ & $4p_{-}$ & $4d$ & $4d_{-}$ &  8.2176(7)b& 	8.2193 [2-4]	& 	8.2206(15)$^{b}$  \\ 
 		     &			    &					&						&		       &					   &					&  8.2155(15)$^{c}$ \\
 33+ [Cu]   & $4d$ & $4d_{+}$ & $4f$ & $4f_{+}$ &  10.2238(7)& 	10.2183 [5-7]	& 	10.2249(15)$^{b}$  \\ 
 		     &			    &					&						&		       &					   &					&  10.2223(15)$^{c}$ \\
 33+ [Cu]   & $4p$ &  $4p_{+}$& $4d$&$4d_{+}$&  10.3555(7) &  10.3588 [3-5]	& 	10.3571(15)$^{b}$  \\ 
 		     &			    &					&						&		       &					   &					&  10.3551(15)$^{c}$ \\
 33+ [Cu]   & $4s$ &  $4s_{+}$& $4p$&$4p_{+}$ &  11.3516(7) &  11.1969 [1-3]	& 	11.3509(15)$^{b}$   \\
 		     &			    &					&						&		       &					   &					&  11.3504(15)$^{c}$ \\
 33+ [Cu]  & $4s$  &  $4s_{+}$& $4p$&$4p_{-}$ &  17.7416(13) & 	17.3523 [1-2]	& 	17.7450$^{d}$  \\
\\
 32+ [Zn] & $4s4d$&  $(4s_{+},4d_{+})_{2}$& $4s4f$ &$(4s_{+},4f_{+})_{3}$ &  9.9127(7) & 	9.8855 [14-30]	& 		    \\
 32+ [Zn] & $4s4p$ &   $(4s_{+},4p_{+})_{1}$& $4s4d$ & $(4s_{+},4d_{+})_{2} $& 10.0851(7)   & 	10.0602 [5-14]	& 		    \\
 32+ [Zn] & $4s^{2}$&  $(4s_{+}^2)_{0}$& $4s4p$ &$(4s_{+},4p_{+})_{1}$ 	&  10.9125(6) & 	10.8361 [1-5]	&  10.911(2)$^{e}$   \\
		     &			    &					&						&		       &					   &					&  10.9101(10)$^{f}$ \\
 32+ [Zn] & $4s4p$ &   $(4s_{+},4p_{-})_{1}$ &  $4p^2$ &$(4p_{-},4p_{+})_{2}$ &  11.2465(8)  & 	11.2393 [3-8]	& 		   \\
 32+ [Zn] & $4s^{2}$ &  $(4s_{+}^2)_{0}$&  $4s4p$& $(4s_{+},4p_{-})_{1}$&  18.8153(19) & 	18.658 [1-3]	& 		  \\
\\
 31+ [Ga] &  $4p$&  $4p_{-}$& $4d$&$4d_{-}$ 						&  7.7461(8)   & 	 7.7150 [1-11]	& 		  \\
 31+ [Ga] &  $4p$ &   $4p_{-}$&  $4s4p_{2}$ & $(4s_{+},(4p_{+}^{2})_{2})_{3/2}$ 	&  8.2146(7)  & 	 8.1951 [1-9]	& 		   \\
 31+ [Ga] &  $4p$& $4p_{+}$&  $4d$& $4d_{-}$ 	&  10.1294(7) & 	 10.0619[2-11]	& 		    \\
 31+ [Ga] &  $4p$&  $4p_{-}$&  $4s4p^{2}$&$(4s_{+},(4p_{+}^{2})_{2})_{5/2}$ 	&  10.8203(7) & 	 10.7701 [1-7]	& 		    \\
 31+ [Ga] &  $4p$&  $4p_{+}$&  $4s4p^{2}$& $(4s_{+},(4p_{+})_{2})_{3/2}$ &  10.9448(20)b& 	 10.8942 [2-9]	&	  \\
 31+ [Ga] &  $4p$&  $4p_{-}$&  $4s4p^{2}$ & $((4s_{+},4p_{-})_{1},4p_{+})_{1/2}$ &  11.1279(7) & 	 11.1055 [1-6]	& 		   \\
 31+ [Ga] &  $4p$&   $4p_{+}$&  $4s4p^{2}$& $((4s_{+},4p_{-})_{1},4p_{+})_{1/2}$ &  16.8114(11) & 	 16.7187 [2-6]	& 		   \\
 31+ [Ga] &  $4p$&  $4p_{-}$&  $4s4p^2$& $4s_{+}$					&  18.3603(16) & 	 18.3159 [1-3]	& 		 \\
 31+ [Ga] &  $4p$&  $4p_{+}$&  $4s4p^{2}$& $((4s_{+},4p_{-})_{1},4p_{+})_{5/2}$ &  19.1476(22)& 	 19.1158 [2-5]	& 		  \\
\\
 30+ [Ge] &  $4p^{2}$&  $(4p_{-}^{2})_{0}$& $4p4d$ &  $(4p_{-},4d_{-})_{1}$ &  7.7020(7)   & 	 7.6492 [1-16]	& 		    \\
 30+ [Ge] &  $4p^{2}$&   $(4p_{-},4p_{+})_{2}$&  $4p4d$&$(4p_{+},4d_{-})_{3}$&  7.9406(7)& 	 7.8988 [3-23]	& 		    \\
 30+ [Ge] &  $4p^{2}$ &   $(4p_{-}^{2})_{0}$&  $4s4p^{3}$& $((4s_{+},4p_{-})_{1},(4p_{+}^{2})_{0})_{1}$ &  8.1809(10)& 	 8.135 [1-13]	& 		    \\
 30+ [Ge] &  $4p^{2}$&   $(4p_{-},4p_{+})_{1}$&  $4p4d$ &$(4p_{-},4d_{-})_{1}$&  9.6994(7)b & 	 9.6032 [2-16]	& 		   \\
 30+ [Ge] &  $4p^{2}$&  $(4p_{-},4p_{+})_{2}$&  $4p4d$&$(4p_{-},4d_{+})_{3}$ &  9.9701(7)& 	 9.9429 [3-15]	& 		   \\
 30+ [Ge] &  $4s4p^{3}$&  $(4s_{+},4p_{+})_{1}$&  $4s4p^{2}4d$&$(4s_{+},4d_{+})_{2}$&  10.0424(7)  & 	 10.0244 [7-33]	& 		  \\
 30+ [Ge] &  $4p^{2}$&   $(4p_{-},4p_{+})_{2}$&  $4s4d$&$(4p_{-},4d_{-})_{2}$&  10.6728(9) & 	 10.6342 [3-14]	& 		    \\
 30+ [Ge] &  $4p^{2}$&   $(4p_{-},4p_{+})_{2}$&  $4s4p^{3}$&$((4s_{+},4p_{-})_{1},(4p_{+}^{2})_{0})_{1}$&  10.7648(8) & 	 10.6784 [3-13]	& 		    \\
 30+ [Ge] &  $4p^{2}$&   $(4p_{-},4p_{+})_{1}$&  $4p4p^{3}$ &$((4s_{+},4p_{-})_{1},(4p_{+}^{2})_{2})_{1}$&  10.9423(7)b & 	 10.8984 [2-12]	& 		   \\
 30+ [Ge] &  $4p^{2}$&  $(4p_{-}^{2})_{0}$&  $4s4p^{3}$ & $(4s_{+},4p_{+})_{1}$&  11.1501(7) & 	 11.1261 [1-7]	& 		    \\
 30+ [Ge] &  $4p^{2}$&   $(4p_{-},4p_{+})_{2}$&  $4p4d$&$(4p_{-},4d_{-})_{2}$&  11.7114(8) & 	 11.7236 [3-10]	& 		   \\
 30+ [Ge] &  $4p^{2}$&  $(4p_{-},4p_{+})_{2}$&  $4s4p^{3}$& $((4s_{+},4p_{-})_{1},(4p_{+}^2)_{2})_{3}$&  12.3270(11)& 	 12.3630 [3-9]	& 		   \\
 30+ [Ge] &  $4p^{2}$&  $(4p_{-},4p_{+})_{2}$&  $4s4p^{3}$& $(4s_{+},4p_{+})_{1}$&  16.5715(9)  & 	 16.5015 [3-7]	& 		  \\
\\
 29+ [As] &  $4p^{3}$&    $4p_{+}$&  $4p^{2}4d$&$((4p_{-},4p_{+})_{2},4d_{-})_{1/2}$ &  7.6812(7)b& 	 7.6310 [1-24]	& 		    \\
 29+ [As] &  $4p^{3}$&    $4p_{+}$&  $4p^{2}4d$&$((4p_{-},4p_{+})_{2},4d_{+})_{5/2}$ &  7.6812(7)b& 	 7.6372 [1-23]	& 		    \\
 29+ [As] &  $4p^{3}$&    $4p_{+}$&  $4p^{2}4d$ &$((4p_{-},4p_{+})_{2},4d_{-})_{3/2}$&  7.8640(8)& 	 7.8150 [1-22]	& 		    \\
 29+ [As] &  $4p^{3}$&    $4p_{+}$&  $4p^{2}4d$ &$((4p_{-},4p_{+})_{1},4d_{+})_{5/2}$&  7.9127(8)& 	 7.8810 [1-21]	& 		    \\
 29+ [As] &  $4p^{3}$ &    $(4p_{-},(4p_{+}^2)_{2})_{3/2}$&  $4p^{2}4d$ &$((4p_{-},4p_{+})_{2},4d_{+})_{5/2}$ &  9.0907(8)    & 	 8.9946 [2-28]	& 		   \\
 29+ [As] &  $4p^{3}$&    $4p_{+}$&  $4p^{2}4d$ & $4d_{+}$&  9.9388(8)& 	 9.9104[1-10]	& 		    \\
 29+ [As] &  $4p^{3}$&  $4p_{+}$&  $4p^{2}4d$&$4d_{-}$ &  10.5983(7)& 	 10.5592[1-9]	& 		   \\
 29+ [As] &  $4p^{3}$&  $4p_{+}$&  $4p^{2}4d$&$4d_{-}$&  11.5407(7)& 	 11.5409[1-7]	& 		   \\
 29+ [As] &  $4p^{3}$&  $4p_{+}$&  $4s4p^{4}$& $(4s_{+},(4p_{+}^{2})_{2})_{5/2}$ &  12.5081(7)& 	 12.5409 [1-6]	& 		    \\
 29+ [As] &  $4p^{3}$&   $(4p_{-},(4p_{+}^2)_{2})_{5/2}$&  $4p^{2}4d$& $4d_{-}$ &  17.3265(12)& 	 17.3085 [3-7]	& 		   \\
 29+ [As] &  $4p^{3}$&   $(4p_{-},(4p_{+}^2)_{0})_{1/2}$&  $4s4p^{4}$& $(4s_{+},(4p_{+}^{2})_{0})_{1/2}$ &  17.6760(14)& 	 17.6538 [4-8]	& 		   \\
 29+ [As] &  $4p^{3}$&  $(4p_{-},(4p_{+}^2)_{2})_{3/2}$&  $4s4p^{4}$&$(4s_{+},(4p_{+}^{2})_{2})_{5/2}$ &  18.4066(17)  & 	 18.4182 [2-6]	& 		   \\
 29+ [As] &  $4p^{3}$&  $(4p_{-},(4p_{+}^2)_{2})_{5/2}$&  $4s4p^{4}$&$(4s_{+},(4p_{+}^{2})_{2})_{5/2}$&  19.6006(24)  & 	 19.6597 [3-6]	& 		   \\
\\
 28+ [Se] &  $4p^{4}$&   $(4p_{+}^{2})_{2}$&  $4p^{3}4d$&$((4p_{-},(4p_{+}^{2})_{2})_{3/2},4d_{-})_{3}$&  7.7090(7)   & 	 7.6446 [1-32]	& 		 \\
 28+ [Se] &  $4p^{4}$&   $(4p_{+}^{2})_{2}$&  $4p^{3}4d$&$((4p_{-},(4p_{+}^2)_{2})_{5/2},4d_{-})_{2}$&  7.7800(7) & 	 7.7198 [1-29]	& 		    \\
 28+ [Se] &  $4p^{4}$&   $(4p_{+}^{2})_{2}$&  $4p^{3}4d$&$(4p_{+},4d_{+})_{3}$&  9.6502(7)b & 	 9.6084 [1-14]	& 		   \\
 28+ [Se] &  $4p^{4}$&   $(4p_{+}^{2})_{0}$&  $4p^{3}4d$&$(4p_{+},4d_{+})_{1}$&  9.8209(13)   & 	 9.7842 [2-15]	& 		   \\
 28+ [Se] &  $4p^{4}$&   $(4p_{+}^{2})_{2}$&  $4p^{3}4d$&$(4p_{+},4d_{+})_{2}$&  9.9790(8)& 	 9.9450 [1-13]	& 		   \\
 28+ [Se] &  $4p^{3}4d$&   $((4p_{+}^{3})_{3/2}, 4d_{+})_{4}$&  $4s4p^{4}4d$&$((4s_{+},(4p_{+}^2)_{0})_{1/2},4d_{+})_{3}$&  10.4393(7) &  10.4893 [12-65]	& 		   \\
 28+ [Se] &  $4p^{4}$&  $(4p_{+}^{2})_{2}$&  $4p^{3}4d$& $(4p_{+},4d_{-})_{3}$&  10.9516(9)  &  10.9310 [1-10]	& 		  \\
 28+ [Se] &  $4p^{4}$&   $(4p_{+}^{2})_{2}$&  $4p^{3}4d$&$(4p_{+},4d_{-})_{1}$&  11.5467(11)  &  11.5440 [1-7]	& 		  \\
 28+ [Se] &  $4p^{4}$&   $(4p_{+}^{2})_{2}$&  $4s4p^{5}$&$(4s_{+},(4p_{+})_{3/2})_{2}$&  12.1372(8)  &  12.1546 [1-6]	& 		   \\
\\
 27+ [Br] &  $4p^{5}$&  $(4p_{+}^3)_{3/2}$&  $4p^{4}4d$ &$((4p_{-},(4p_{+}^3)_{3/2})_{2},4d_{-})_{5/2}$ &  7.6242(7)    & 	 7.5549 [1-27]	& 		   \\
 27+ [Br] &  $4p^{5}$&  $(4p_{+}^3)_{3/2}$&  $4p^{4}4d$&$(4p_{-},(4p_{+}^3)_{2},4d_{-})_{1/2}$ &  7.7321(12)   & 	 7.6606 [1-25]	& 		   \\
 27+ [Br] &  $4p^{5}$&  $(4p_{+}^3)_{3/2}$&  $4p^{4}4d$&$((4p_{+}^2)_{2},4d_{+}{_{5/2}})_{5/2}$ &  9.5321(7)   & 	 9.4986 [1-14]	& 		 \\
 27+ [Br] &  $4p^{5}$&  $(4p_{+}^3)_{3/2}$&  $4s$&$4s_{+}$ &  9.6479(7)b   & 	9.6142 [1-13]	& 		 \\
 27+ [Br] &  $4p^{5}$&  $(4p_{+}^3)_{3/2}$&  $4p^{4}4d$&$((4p_{+}^2)_{0},4d_{+}{_{5/2}})_{5/2}$&  9.8126(9)    & 	 9.7848 [1-12]	& 		  \\
 27+ [Br] &  $4p^{5}$&  $(4p_{+}^3)_{3/2}$&  $4p^{4}4d$&$((4p_{+}^2)_{2},4d_{+}{_{5/2}})_{1/2}$&  9.9000(7)   &	 9.8682 [1-11]	& 		   \\
\\
 26+ [Kr] &  $4p^{6}$&  $(4p_{+}^{4})_{0}$&   $4p^{5}4d$&$(4p_{-},4d_{-})_{1}$  &  7.5901(7)   & 	 7.4940 [1-13]	& 		   \\
 26+ [Kr] &  $4p^{6}$&   $(4p_{+}^{4})_{0}$&   $4p^{5}4d$&$((4p_{+}^{3})_{3/2},4d_{+})_{1}$&  9.4330(7)   & 	 9.3654 [1-9]	& 		  \\
 26+ [Kr] &  $4p^{6}$&   $(4p_{+}^{4})_{0}$&   $4p^{5}4d$&$((4p_{+}^{3})_{3/2},4d_{-})_{1}$&  11.2694(9)   & 	 11.2303 [1-3]	& 		  \\
\hline
\multicolumn{8}{l}{ $^{a}$ - \cite{Daido:99}, $^{b}$ - \cite{1402-4896-24-4-008}, $^{c}$ -\cite{Doschek:88}, $^{d}$ -  \cite{PhysRevA.38.288}, $^{e}$ -  \cite{PhysRevLett.45.609}, $^{f}$ -\cite{Acquista:84} }

\end{longtable}

\newpage

\begin{longtable}{l l l l l l l l l}
\caption{Wavelengths (nm) of highly charged erbium. 
The numbers in parentheses are the wavelength uncertainties in units of 
the last significant digit. The numbers in square
brackets in column ``FAC" are the ordinal numbers for the 
lower and upper levels. 
b-blended line. (*2)-wavelength derived from second order measurement.
\label{t:full_results_er}}\\
 Stage &  \multicolumn{2}{l}{ Lower Level} &   \multicolumn{2}{l}{ Upper Level}  &  Wavelength &  FAC &  Prev. Exp.  \\
&  Conf. &  State &  Conf.  &  State &    &    &     \\ 
\hline
 40+ [Ni] &  $3d^{9}4p$&  $((3d_{+}^{5})_{5/2},4p_{+})_{4}$&  $3d^{9}4d$ &  $((3d_{+}^{5})_{5/2},4d_{+})_{5}$ &  8.7119(12)   & 	8.7307 [10-23]	& 		&    \\ 
 40+ [Ni] &  $3d^{9}4s$ &  $((3d_{+}^{5})_{5/2},4s_{+})_{3}$&  $3d^{9}4p$ &  $((3d_{+}^5)_{5/2},4p_{-})_{3}$ &  14.8490(14)    & 	14.8038 [2-7]	& 		&    \\ 
 40+ [Ni] &  $3d^{9}4s$ &  $((3d_{+}^{5})_{5/2},4s_{+})_{3}$&  $3d^{9}4p$ &  $((3d_{+}^5)_{5/2},4p_{-})_{2}$  &  15.0340(12)   & 	15.0153 [2-6]	& 		&    \\ 
 40+ [Ni] &  $3d^{9}4s$ &  $((3d_{+}^5)_{5/2},4s_{+})_{2}$&  $3d^{9}4p$ &  $((3d_{+}^5)_{5/2},4p_{-})_{3}$ &  15.1355(13)  & 	15.1163 [3-7]	& 		&    \\ 
 40+ [Ni] &  $3d^{9}4s$ &   $((3d_{-}^{3})_{3/2},4s_{+})_{2}$&  $3d^{9}4p$ &  $((3d_{-}^{3})_{3/2},4p_{-})_{2}$&  15.2200(12)  & 	15.2079 [5-8]	& 		&    \\ 
 40+ [Ni] &  $3d^{9}4s$ &  $((3d_{+}^5)_{5/2},4s_{+})_{2}$&  $3d^{9}4p$ &  $((3d_{+}^5)_{5/2},4p_{-})_{2}$&  15.3297(12)   & 	15.3369 [3-6]	& 		&    \\ 
\\
 39+ [Cu] &  $4p$&   $4p_{-}$ & 4d& $4d_{-}$  &  6.3391(14)   & 	 6.3379 [2-4]	& 	6.3403(15)$^{a}$  \\ 
 39+ [Cu] &  $4s$ &   $4s_{+}$&  $4p$& $4p_{+}$&  8.3796(12)   & 	 8.3654 [1-3]	& 	8.3813(15)$^{a}$ \\
 39+ [Cu] &  $4p$&   $4p_{+}$&  $4d$& $4d_{+}$&  8.5733(12)   & 	 8.5712 [3-5]	& 	8.5760(15)$^{a}$  \\
 39+ [Cu] &  $4s$&  $4s_{+}$&  $4p$& $4p_{-}$ &  14.8817(13)  & 	14.8305 [1-2]	& 	  \\
\\
 38+ [Zn]  &  $4s4p$&  $(4s_{+},4p_{-})_{1}$&  $4s4d$& $(4s_{+},4d_{-})_{2}$ &  6.2733(12) & 	6.2686 [3-10]	& 		&  \\
 38+ [Zn]  &  $4s^2$&   $(4s_{+}^2)_{0}$&  $4s4p$& $(4s_{+},4p_{+})_{1}$&  8.1312(12) & 	8.0900 [1-5]	& 	8.131(2)$^{b}$ \\
 38+ [Zn]  &  $4s4p$&  $(4s_{+},4p_{+})_{1}$&  $4s4d$& $(4s_{+},4d_{+})_{2}$&  8.1732(21)  & 	8.1548 [5-14]	& 		  \\
 38+ [Zn]  &  $4s^2$&   $(4s_{+}^2)_{0}$&  $4s4p$& $(4s_{+},4p_{-})_{1}$&  15.6499(12) & 	15.5430 [1-3]	& 	   \\
\\
 37+ [Ga] &  $4p$ &  $4p_{-}$&  $4d$& $4d_{-}$&  6.1755(12)     &  6.1626 [1-9]	& 		 \\
 37+ [Ga] &  $4p$ &  $4p_{-}$&  $4s4p^2$ & $((4s_{+},4p_{-})_{1},4p_{+})_{1/2}$&  8.0777(14)    &  8.0497 [1-7]	& 		\\
 37+ [Ga] &  $4p$	& $4p_{-}$	& $4s4p^2$ &  $((4s_{+},4p_{-})_{1},4p_{+})_{3/2}$	&  8.2227(12)b & 8.2109 [1-6]	&  	 \\
 37+ [Ga] &  $4p$ &   $4p_{-}$&  $4s4p^2$ & $((4s_{+},4p_{-})_{0},4p_{+})_{3/2}$ &  9.3102(12)   &  9.3281 [1-4]	& 		 \\
 37+ [Ga] &  $4p$ &  $4p_{+}$&  $4s4p^2$ & $((4s_{+},4p_{-})_{2},4p_{+})_{1/2}$&  13.5942(13)    &  13.4904 [2-7]	& 		\\
 37+ [Ga] &  $4p$ &  $4p_{+}$&  $4s4p^2$ & $((4s_{+},4p_{-})_{1},4p_{+})_{3/2}$&  14.0226(13)&  13.9493 [2-6]	& 		  \\
 37+ [Ga] &  $4p$ &  $4p_{-}$&  $4s4p^2$ & $4s_{+}$&  15.1604(12) &  15.1163 [1-3]	& 		\\
 37+ [Ga] &  $4p$ &  $4p_{+}$&  $4s4p^2$ & $((4s_{+},4p_{-})_{1},4p_{+})_{5/2}$ &  15.8957(13)&  15.8643 [2-5]	& 		 \\
\\
 36+ [Ge]    &  $4p^2$&  $(4p_{-}^{2})_{0}$&  $4p4d$& $(4p_{-},4d_{-})_{1}$ &  6.1041(12)&  6.0786 [1-11]	& 		   \\
 36+ [Ge]	&  $4p^2$	&  $(4p_{-}^{2})_{0}$ &  $4s4p^3$ &  	$(4s_{+},4p_{+})_{1}$	&	 8.2214(12)b &  8.2014 [1-7] &  \\
 36+ [Ge]    &  $4p^2$&  $(4p_{-},4p_{+})_{2}$&  $4s4p^3$& $(4s_{+},4p_{+})_{1}$ &  13.7334(12) &  13.6693 [3-7]	& 		   \\
\\
 35+ [As]  &  $4p^3$&  $4p_{+}$&  $4p^{2}4d$ & $((4p_{-},4p_{+})_{2},4d_{-})_{3/2}$ &  6.0497(13)&  6.0137 [1-21]	& 		   \\
 35+ [As]  &  $4p^3$&  $4p_{+}$&  $4p^{2}4d$ & $((4p_{-},4p_{+})_{1},4d_{-})_{5/2}$&  6.0727(23) &  6.0435 [1-20]	& 		   \\
 35+ [As]  &  $4p^3$&   $(4p_{-},(4p_{+})_{0})_{1/2}$&  $4p^{2}4d$ & $((4p_{-},4p_{+})_{1},4d_{+})_{7/2}$&  8.8392(13)&  8.8433 [3-17]	& 		   \\
 35+ [As]  &  $4p^3$&   $(4p_{-},(4p_{+}^{2})_{2})_{3/2}$&  $4s4p^{4}$ & $(4s_{+},(4p_{+}^{2})_{2})_{5/2}$ &  15.4209(13)&  15.4358 [2-7]	& 		   \\
 35+ [As]  &  $4p^3$&   $(4p_{-},(4p_{+}^{2})_{2})_{5/2}$&  $4s4p^{4}$& $(4s_{+},(4p_{+}^{2})_{2})_{5/2}$&  16.3966(13)&  16.4483 [3-7]	& 		   \\
 35+ [As]  &  $4p^3$&  $(4p_{-},(4p_{+}^{2})_{2})_{5/2}$&  $4p^{2}4d$& $4d_{-}$&  17.1697(15)&  17.1498 [3-6]	& 		 \\
\\
 34+ [Se]  &  $4p^4$&  $(4p_{+}^2)_{2}$ &   $4p^{3}4d$&  $((4p_{-},(4p_{+}^{2})_{2})_{3/2},4d_{-})_{3}$ &  5.9929(18) &  5.9580 [1-30]	& 		 \\
 34+ [Se]  &  $4p^4$&  $(4p_{+}^2)_{2}$&   $4p^{3}4d$ &  $((4p_{-},(4p_{+}^{2})_{2})_{5/2},4d_{-})_{2}$ &  6.0230(22)&  5.9797 [1-29]	& 		 \\
 34+ [Se]  &  $4p^4$&  $(4p_{+}^2)_{2}$&   $4p^{3}4d$ &  $(4p_{+},4d_{+})_{3}$&  7.8933(12) &  7.8593 [1-14]	& 		 \\
 34+ [Se]  &  $4p^4$&   $(4p_{+}^{2})_{0}$&   $4p^{3}4d$&  $(4p_{+},4d_{+})_{1}$&  7.9546(12) &  7.9199 [2-15]	& 		 \\
 34+ [Se]  &  $4p^4$&  $(4p_{+}^2)_{2}$&   $4p^{3}4d$&  $(4p_{+},4d_{+})_{2}$&  8.0042(12)  &  7.9737 [1-13]	& 		 \\
 34+ [Se]  &  $4p^4$&  $(4p_{+}^2)_{2}$&   $4p^{3}4d$&  $(4p_{+},4d_{-})_{3}$ &  9.1787(15) &  9.169 [1-9]	& 		 \\
\\
 33+ [Br]   &  $4p^5$&  $(4p_{+}^{3})_{3/2}$&  $4p^{4}4d$ &  $((4p_{-},(4p_{+}^{3})_{3/2})_{2},4d_{-})_{5/2}$ &  5.9479(6)(*2)&  5.9049 [1-26] & 		 \\ 
 33+ [Br]   &  $4p^5$&  $(4p_{+}^{3})_{3/2}$&  $4p^{4}4d$ &  $((4p_{-},(4p_{+}^{3})_{3/2})_{2},4d_{-})_{3/2}$&  5.9789(7)b(*2)&  5.9390 [1-25] & 		 \\ 
 33+ [Br]   &  $4p^5$&  $(4p_{+}^{3})_{3/2}$&  $4p^{4}4d$ &  $((4p_{-},(4p_{+}^{3})_{3/2})_{2},4d_{-})_{1/2}$&  5.9789(7)b(*2)&  	5.9443 [1-24] & 		 \\ 
 33+ [Br]   &  $4p^5$&  $(4p_{+}^{3})_{3/2}$&  $4p^{4}4d$ &  $((4p_{+}^{2})_{2},4d_{+})_{5/2}$&  7.7630(12)&  7.7217 [1-14]	& 		 \\
 33+ [Br]   &  $4p^5$&  $(4p_{+}^{3})_{3/2}$&  $4s$ & $4s_{+}$& 7.8578(10) &  7.8181[1-13]	& \tiny		 \\ 
 33+ [Br]   &  $4p^5$&   $(4p_{+}^{3})_{3/2}$&  $4p^{4}4d$&  $((4p_{+}^{2})_{2},4d_{+})_{3/2}$&  7.8717(13)&  7.8352 [1-12]	& 		 \\
 33+ [Br]   &  $4p^5$&   $(4p_{+}^{3})_{3/2}$&  $4p^{4}4d$&  $((4p_{+}^{2})_{2},4d_{+})_{1/2}$&  9.2830(14)&  9.2630 [1-6]	& 		 \\
 33+ [Br]   &  $4p^5$&   $(4p_{+}^{3})_{3/2}$&  $4p^{4}4d$&  $((4p_{+}^{2})_{2},4d_{-})_{5/2}$&  9.4626(47)&  9.4519 [1-5]	& 		 \\
 33+ [Br]   &  $4p^5$&   $(4p_{+}^{3})_{3/2}$&  $4p^{4}4d$&  $((4p_{+}^{2})_{2},4d_{-})_{3/2}$&  9.5823(12)&  9.5710 [1-3]	& 		 \\
\\
 32+ [Kr]    &  $4p^6$&   $(4p_{+}^{4})_{0}$&  $4p^{5}4d$ & $(4p_{-},4d_{-})_{1}$ &  5.9356(13)&  5.8792 [1-13]	& 		 \\
 32+ [Kr]    &  $4p^6$&   $(4p_{+}^{4})_{0}$&  $4p^{5}4d$& $((4p_{+}^{3})_{3/2},4d_{+})_{1}$ &  7.6746(12)&  7.6221 [1-9]	& 		 \\
 32+ [Kr]    &  $4p^6$&   $(4p_{+}^{4})_{0}$&  $4p^{5}4d$& $((4p_{+}^{3})_{3/2},4d_{-})_{1}$&  9.4129(12)&  9.3900 [1-3]		& 		 \\
\\
 31+ [Rb]    &  $4d$&  $4d_{-}$&  $4p^{5}4d^{2}$ & $(4p_{-}, (4d_{-}^{2})_{2})_{3/2}$&  5.8139(13)&  5.7583 [1-45]	& 		 \\
 31+ [Rb]    &  $4d$&  $4d_{-}$&  $4p^{5}4d^{2}$& $(4p_{-},(4d_{-}^{2})_{0})_{1/2}$ &  5.9709(18)&  5.9212 [1-40]	& 		 \\
 31+ [Rb]    &  $4d$&  $4d_{-}$&  $4p^{5}4d^{2}$& $(4p_{-},(4d_{-}^{2})_{2})_{5/2}$ &  6.2905(12)&  6.2377 [1-34]	& 		 \\
 31+ [Rb]    &  $4d$&  $4d_{-}$&  $4f$& $4f_{-}$ &  6.9150(11) & 6.8765[1-32]	& 		 \\
 31+ [Rb]    &  $4d$&  $4d_{-}$&  $4p^{5}4d^{2}$& $(((4p_{+}^{3})_{3/2},4d_{-})_{3},4d_{+})_{1/2}$ &  7.4543(12)&  7.3946 [1-28]	& 		 \\
 31+ [Rb]    &  $4d$&  $4d_{-}$&  $4p^{5}4d^{2}$& $(((4p_{+})_{3/2},4d_{-})_{3},4d_{+})_{3/2}$ &  7.8144(12)&  7.7673 [1-25]	& 		 \\
 31+ [Rb]    &  $4d$&  $4d_{-}$&  $4p^{5}4d^{2}$& $((4p_{+}^{3})_{3/2},(4d_{+}^{2})_{4})_{5/2}$ &  7.9633(13)b&  7.9128 [1-21]	& 		 \\
 31+ [Rb]    &  $4d$&  $4d_{-}$&  $4p^{5}4d^{2}$& $((4p_{+}^{3})_{3/2},(4d_{+}^{2})_{2})_{5/2}$ &  7.9633(13)b&  7.9255 [1-22]	& 		 \\
 31+ [Rb]    &  $4d$&  $4d_{-}$&  $4p^{5}4d^{2}$& $(((4p_{+}^{3})_{3/2},4d_{-})_{3},4d_{+})_{5/2}$&  8.2514(12)&  8.2165 [1-17]	& 		 \\
 31+ [Rb]    &  $4d$&  $4d_{-}$&  $4p^{5}4d^{2}$& $(((4p_{+}^{3})_{3/2}, 4d_{-})_{1}, 4d_{+})_{3/2}$&  8.6289(12)&  8.6045 [1-11]	& 		 \\
 31+ [Rb]    &  $4d$&  $4d_{-}$&  $4p^{5}4d^{2}$& $((4p_{+}^{3})_{3/2},(4d_{-}^{2})_{0})_{3/2}$ &  8.8905(13)&  8.8592 [1-8]	& 		 \\
 31+ [Rb]    &  $4d$&  $4d_{+}$&  $4p^{5}4d^{2}$& $(((4p_{+}^{3})_{3/2}, 4d_{-})_{2}, 4d_{+})_{7/2}$&  8.9963(12) &  8.9544 [2-16]	& 		 \\
 31+ [Rb]    &  $4d$&  $4d_{+}$&  $4p^{5}4d^{2}$& $(((4p_{3}^3)_{3/2}, 4d_{-})_{3}, 4d_{+})_{7/2}$ &  9.2249(12)&  9.1879 [2-14]	& 		 \\
 31+ [Rb]    &  $4d$&  $4d_{-}$&  $4p^{5}4d^{2}$& $(((4p_{+}^{3})_{3/2}, (4d_{-}^{2})_{2})_{5/2}$&  9.3670(13)&  9.3340 [1-5]	& 		 \\
 31+ [Rb]    &  $4d$&  $4d_{+}$&  $4p^{5}4d^{2}$& $((4p_{+}^{3})_{3/2}, (4d_{-}^{2})_{2})_{7/2}$ &  10.1342(12)  &  10.0913 [2-6]	& 		 \\
\hline
\multicolumn{8}{l}{ $^{a}$ -  \cite{1402-4896-24-4-008}, $^{b}$ -  \cite{Acquista:84},  }

\end{longtable}

\newpage

\begin{longtable}{l l l r r l r}
\caption{Energy levels of highly charged samarium. 
RMBPT -relativistic many body perturbation theory \cite{Safronova200647}, FAC - flexible atomic code \cite{FAC}, $\dagger$-weighted average from two transitions, $\dagger\dagger$-weighted average from three transitions  \label{t:full_results_sm_energy}}\\
 Stage and  &  Configuration &  State &  Level number&  Energy  & &  Unc. \\
 sequence &                &        & (FAC)  & (cm${^{-1}}$)              & &  (cm${^{-1}}$) \\
\hline
 34+ [Ni] &   $3d^{10}$&  $(3d_{+}^{6})_{0}$ 				&  1 &  0	 & &    0 \\  
 34+ [Ni] &   $3d^{9}4s$&  $((3d_{+}^{5})_{5/2},4s_{+})_{3}$ 	&  2 &   7 522 190 &+x	&  RMBPT \\  
 34+ [Ni] &  $3d^{9}4s$ &  $((3d_{+}^{5})_{5/2},4s_{+})_{2}$ 	&  3 &  7 533 150 &+x		& 	50$\dagger$ 	 \\ 
 34+ [Ni] &  $3d^{9}4s$ &  $((3d_{-}^{3})_{3/2},4s_{+})_{2}$ 	&  5  & 	  7 752 539 &+y	&  RMBPT		 \\
 34+ [Ni] &  $3d^{9}4p$ &  $((3d_{+}^5)_{5/2},4p_{-})_{2}$ 	&  6 & 	  8 081 080 &+x	&  	40	 \\ 
 34+ [Ni] &  $3d^{9}4p$ &  $((3d_{+}^5)_{5/2},4p_{-})_{3}$ 	&  7 & 	 8 089 190 &+x	&  	50	 \\
 34+ [Ni] &  $3d^{9}4p$ & $((3d_{-}^3)_{3/2},4p_{-})_{2}$ 	&  8 & 	8 304 770 &+y & 	60	 \\
 34+ [Ni] &  $3d^{9}4p$ &  $((3d_{-}^{3})_{3/2},4p_{-})_{1}$ 	&  9 & 	 8 322 930 &+z	&  RMBPT		 \\
 34+ [Ni] &  $3d^{9}4p$ &  $((3d_{+}^5)_{5/2},4p_{+})_{4}$ 	&  10 & 	8 404 586	&+a &  FAC		 \\ 
 34+ [Ni] &  $3d^{9}4p$ &  $((3d_{+}^5)_{5/2},4p_{+})_{1}$ 	&  12 & 	8 423 520&+z	&  200		 \\
 34+ [Ni] &  $3d^{9}4d$ &  $((3d_{+}^5)_{5/2},4d_{+})_{5}$ 	&  23 & 	9 357 800 &+a & 	100	 \\
 34+ [Ni] &  $3d^{9}4d$ &  $((3d_{-}^3)_{3/2},4d_{-})_{0}$ 	&  35  & 	9 782 910&+z 	& 	150 \\
\\
 33+ [Cu] &   $4s$&  $4s_{+}$ &  1& 0	 &&    0 \\  
 33+ [Cu] &   $4p$&  $4p_{-}$ &  2& 563 650&	 &   40  \\  
 33+ [Cu] &   $4p$&  $4p_{+}$ &  3& 880 940&	 &  50    \\  
 33+ [Cu] &   $4d$&  $4d_{-}$ &  4& 1 780 550 &  &  110    \\  
 33+ [Cu] &   $4d$&  $4d_{+}$ &  5& 1 846 610&	 &  80   \\  
 33+ [Cu] &   $4f$&  $4f_{+}$  &  7& 2 824 720	 &&   110  \\  
\\
 32+ [Zn] &   $4s^{2}$&  $(4s_{+}^2)_{0}$ &  1 &  0	 &&    0 \\  
 32+ [Zn] &   $4s4p$&  $(4s_{+},4p_{-})_{1}$ &  3&  531 480&	 &  50  \\  
 32+ [Zn] &   $4s4p$&  $(4s_{+},4p_{+})_{1}$ &  5 & 916 380&	 &  50  \\  
 32+ [Zn] &   $4p^{2}$&  $(4p_{-},4p_{-})_{2}$ &  8 &  1 420 650&	 &  80  \\  
 32+ [Zn] &   $4s4d$&  $(4s_{+},4d_{+})_{2}$ &  14 &  1 907 940&	 &  90  \\  
 32+ [Zn] &   $4s4f$&  $(4s_{+},4f_{+})_{3}$ &  30 &  2 916 740	 &&   110 \\  
\\
 31+ [Ga] &   $4p$&  $4p_{-}$ &  1 &  0	& &  0 \\  
 31+ [Ga] &   $4p$&  $4p_{+}$ &  2 &   303 790	& &  50$\dagger\dagger$  \\  
 31+ [Ga] &   $4s4p^{2}$&  $4s_{+}$ & 3 &  544 650	 &&  50 \\  
 31+ [Ga] &   $4s4p^{2}$&  $((4s_{+},4p_{-})_{1},4p_{+})_{5/2}$ &  5 &  	826 040& &  80 \\  
 31+ [Ga] &   $4s4p^{2}$&  $((4s_{+},4p_{-})_{1},4p_{+})_{3/2}$ &  6 & 	898 640& &  50 \\  
 31+ [Ga] &   $4s4p^{2}$&  $((4s_{+},4p_{-})_{1},4p_{+})_{1/2}$ &  7 & 	924 190& &   60 \\  
 31+ [Ga] &   $4s4p^{2}$&  $(4s_{+},(4p_{+}^{2})_{2})_{3/2}$ &  9 & 1 217 350	& &  100 \\  
 31+ [Ga] &   $4d$&  $4d_{-}$ &  11 &  1 290 970&	 &  130 \\  
\\
 30+ [Ge] &   $4p^{2}$&  $(4p_{-}^2)_{0}$ &  1&   0	& &  0 \\  
 30+ [Ge] &   $4p^{2}$&  $(4p_{-},4p_{+})_{1}$ &  2 &  267 370&	 &  140  \\  
 30+ [Ge] &   $4p^{2}$&  $(4p_{-},4p_{+})_{2}$ &  3 &  	293 400& &  60$\dagger$ \\  
 30+ [Ge] &   $4s4p^{3}$&  $(4s_{+},4p_{+})_{1}$ &  7 &  896 850&	 &  50 \\  
 30+ [Ge] &   $4s4p^{3}$&  $((4s_{+},4p_{-})_{1},(4p_{+}^{2})_{2})_{3}$ &  9 &  1 104 630&	 &  90 \\  
 30+ [Ge] &   $4p4d$&  $(4p_{-},4d_{-})_{2}$ &  10 &   1 147 270 &&  80 \\  
 30+ [Ge] &   $4p4p^{3}$&  $((4s_{+},4p_{-})_{1},(4p_{+}^{2})_{2})_{1}$ &  12 &   1 181 260 &&  160 \\  
 30+ [Ge] &   $4s4p^{3}$&  $((4s_{+},4p_{-})_{1},(4p_{+}^{2})_{0})_{1}$ &  13 &  1 222 350&	 &  160 \\  
 30+ [Ge] &   $4p4d$&  $(4p_{-},4d_{+})_{2}$ &  14 &  1 230 370 & &   100\\  
 30+ [Ge] &   $4p4d$&  $(4p_{-},4d_{+})_{3}$ &  15 &  1 296 400 & &   90\\  
 30+ [Ge] &   $4p4d$&  $(4p_{-},4d_{-})_{1}$ &  16 &  1 298 360	& &  120 \\  
 30+ [Ge] &   $4p4d$&  $(4p_{+},4d_{-})_{3}$ &  23 & 1 552 740& &  130\\  
 30+ [Ge] &   $4s4p^{2}4d$&  $(4s_{+},4d_{+})_{2}$ &  33&  1 892 630	& &  900 \\  
\\
 29+ [As] &   $4p^{3}$&  $(4p_{+}^3)_{3/2}$ &  1 &  0	 &&  0 \\  
 29+ [As] &   $4p^{3}$&  $(4p_{-},(4p_{+}^2)_{2})_{3/2}$ &  2 &  256 210&	 &  70 \\  
 29+ [As] &   $4p^{3}$&  $(4p_{-},(4p_{+}^2)_{2})_{5/2}$ &  3&  289 330 & &  50$\dagger$   \\  
 29+ [As] &   $4p^{3}$&  $(4p_{-},(4p_{+}^2)_{0})_{1/2}$ &  4 &  331 860 &+x	 &  FAC \\  
 29+ [As] &   $4s4p^{4}$&  $(4s_{+},(4p_{+}^2)_{2})_{5/2}$ &  6 &  799 490&	 &  50 \\  
 29+ [As] &   $4p^{2}4d$&  $4d_{-}$ &  7 &  	866 500 &&  50 \\  
 29+ [As] &   $4s4p^{3}$&  $(4s_{+},(4p_{+}^{2})_{0})_{1/2}$ &  8 &  	897 600 &+x &  50 \\  
 29+ [As] &   $4p^{2}4d$&  $4d_{-}$ &  9 & 	943 540 &&  60 \\  
 29+ [As] &   $4p^{2}4d$&  $4d_{+}$ &  10 &  	1 006 160 &&  100 \\  
 29+ [As] &   $4p^{2}4d$&  $((4p_{-},4p_{+})_{1},4d_{+})_{5/2}$ &  21 &  	1 263 780 &&  120 \\  
 29+ [As] &   $4p^{2}4d$&  $((4p_{-},4p_{+})_{2},4d_{-})_{3/2}$ &  22 &  	1 271 620 &&  120 \\  
 29+ [As] &   $4p^{2}4d$&  $((4p_{-},4p_{+})_{2},4d_{+})_{5/2}$ &  23 &  	1 301 880b &&  120 \\  
 29+ [As] &   $4p^{2}4d$&  $((4p_{-},4p_{+})_{2},4d_{-})_{1/2}$ &  24 &  	1 301 880b &&  120 \\  
 29+ [As] &   $4p^{2}4d$&  $((4p_{-},(4p_{+})_{2}),4d_{+})_{5/2}$ &  28&  1 356 240	& &  120  \\  
\\
 28+ [Se] &   $4p^{4}$&  $(4p_{+}^2)_{2}$ &  1 &  0	 &&  0 \\  
 28+ [Se] &   $4p^{4}$&  $(4p_{+}^2)_{0}$ &  2 &  	53 669 &+x &  FAC \\  
 28+ [Se] &   $4s4p^{5}$&  $(4s_{+}, (4p_{+}^{3})_{3/2})_{2}$ &  6&  823 910 &&  50 \\  
 28+ [Se] &   $4p^{3}4d$&  $(4p_{+},4d_{-})_{1}$ &  7 &  866 050	& &  80 \\  
 28+ [Se] &   $4p^{3}4d$&  $(4p_{+},4d_{-})_{3}$ &  10 &  913 110	& &  70 \\  
 28+ [Se] &   $4p^{3}4d$&  $(4p_{+},4d_{+})_{4}$ &  12 &  964 703& +y	 &   FAC \\  
 28+ [Se] &   $4p^{3}4d$&  $(4p_{+},4d_{+})_{2}$ &  13 &  1 002 100&	 &  80 \\  
 28+ [Se] &   $4p^{3}4d$&  $(4p_{+},4d_{+})_{3}$ &  14 &  1 036 250&	 &  70  \\  
 28+ [Se] &   $4p^{3}4d$&  $(4p_{+},4d_{+})_{1}$ &  15 &   1 071 900 &+x &  130 \\  
 28+ [Se] &   $4p^{3}4d$&  $((4p_{-},(4p_{+}^{2})_{2})_{5/2},4d_{-})_{2}$ &  29 &  1 285 350&	 &  120  \\  
 28+ [Se] &   $4p^{3}4d$&  $((4p_{-},(4p_{+}^{2})_{2})_{3/2},4d_{-})_{3}$ &  32 &  1 297 190&	 &  130  \\  
 28+ [Se] &   $4s4p^{4}4d$&  $((4s_{+},(4p_{+}^{2})_{0})_{1/2},4d_{+})_{3}$ &  65 &  1 922 620& +y &   70\\  
\\
 27+ [Br] &   $4p^{5}$&  $(4p_{+}^{3})_{3/2}$ &  1 &  0	& &  0 \\  
 27+ [Br] &   $4p^{4}4d$&  $((4p_{+}^{2})_{2},4d_{+})_{1/2}$ &  11 &  1 010 100&	 &  70  \\  
 27+ [Br] &   $4p^{4}4d$&  $((4p_{+}^{2})_{0},4d_{+})_{5/2}$ &  12 &  1 019 100&	 &  100  \\  
 27+ [Br] &   $4s$&  $4s_{+}$ &  13 & 1 036 490 &  	 &  70  \\  
 27+ [Br] &   $4p^{4}4d$&  $((4p_{+}^{2})_{2},4d_{+})_{5/2}$ &  14 &  1 049 090&	 &  80  \\  
 27+ [Br] &   $4p^{4}4d$&  $((4p_{-},(4p_{+}^{3})_{3/2})_{2},4d_{-})_{1/2}$ &  25 &  1 293 300	& &  200 \\  
 27+ [Br] &   $4p^{4}4d$&  $((4p_{-},(4p_{+}^{3})_{3/2})_{2},4d_{-})_{5/2}$ &  27 &  1 311 620& &  120  \\  
\\
 26+ [Kr] &   $4p^{6}$&  $(4p_{+}^{4})_{0}$ &  1 &  0	 &&  0 \\  
 26+ [Kr] &   $4p^{5}4d$&  $((4p_{+}^{3})_{3/2},4d_{-})_{1}$ &  3 &  887 360&	 &   70\\  
 26+ [Kr] &   $4p^{5}4d$&  $((4p_{+}^{3})_{3/2},4d_{+})_{1}$ &  9 &  1 060 100	& &   80\\  
 26+ [Kr] &   $4p^{5}4d$&  $(4p_{-},4d_{-})_{1}$ &  13 & 	1 317 500 &&  120 \\  

\end{longtable}

\newpage

\begin{longtable}{l l l r r l r}
\caption{Energy levels of highly charged erbium. RMBPT -relativistic many-body perturbation theory \cite{Safronova200647}, FAC - flexible atomic code \cite{FAC}, $\dagger$-weighted average from two transitions \label{t:full_results_er_energy}}\\
 Stage and  &  Configuration &  State &  Level number&  Energy  & &  Unc. \\
 sequence &                &        & (FAC)  & (cm${^{-1}}$)              & &  (cm${^{-1}}$) \\
\hline
 40+ [Ni] &   $3d^{10}$&  $(3d_{+}^{6})_{0}$ &  1 &  0	& &    0 \\  
 40+ [Ni] &   $3d^{9}4s$&  $((3d_{+}^{5})_{5/2},4s_{+})_{3}$ &  2&    9 931 473 &+x	&  RMBPT \\  
 40+ [Ni] &   $3d^{9}4s$&  $((3d_{+}^{5})_{5/2},4s_{+})_{2}$ &  3 &   9 944 260	&+x&  60$\dagger$ \\  
 40+ [Ni] &   $3d^{9}4s$&  $((3d_{-}^{3})_{3/2},4s_{+})_{2}$ &  5&   10 294 420& +y	&  RMBPT \\  
 40+ [Ni] &   $3d^{9}4p$&  $((3d_{+}^{5})_{5/2},4p_{-})_{2}$ &  6 &   10 596 630 &+x	&  60 \\  
 40+ [Ni] &   $3d^{9}4p$&  $((3d_{+}^{5})_{5/2},4p_{-})_{3}$ &  7 &   10 604 920 &+x	&  70 \\  
 40+ [Ni] &   $3d^{9}4p$&  $((3d_{-}^{3})_{3/2},4p_{-})_{2}$ &  8 &   10 951 450 &+y	&   50\\  
 40+ [Ni] &   $3d^{9}4p$&  $((3d_{+}^{5})_{5/2},4p_{+})_{4}$ &  10 &   11 127 857 &+z &   FAC\\  
 40+ [Ni] &   $3d^{9}4d$&  $((3d_{+}^{5})_{5/2},4d_{+})_{5}$ &   23 &  12 275 710 &+z	&  160 \\  
\\
 39+ [Cu] &   $4s$&  $4s_{+}$ &  1 & 0	& &  0   \\  
 39+ [Cu] &   $4p$&  $4p_{-}$ &  2 &   671 970	 &&   60  \\  
 39+ [Cu] &   $4p$&  $4p_{+}$ &  3 &  1 193 370	& &  170   \\  
 39+ [Cu] &   $4d$&  $4d_{-}$ &  4 &  2 249 480	 &&   360  \\  
 39+ [Cu] &   $4d$&  $4d_{+}$ &  5 &  2 359 780	& &   230  \\  
\\
 38+ [Zn] &   $4s^{2}$&  $(4s_{+}^2)_{0}$ &  1 & 0	& &  0   \\  
 38+ [Zn] &   $4s4p$&  $(4s_{+},4p_{-})_{1} $ &  3 &  638 980&	 &   50  \\  
 38+ [Zn] &   $4s4p$&  $(4s_{+},4p_{+})_{1} $ & 5 & 1 229 820	& &  180   \\  
 38+ [Zn] &   $4s4d$&  $(4s_{+},4d_{-})_{2} $ & 10 &   2 233 040	& &   310  \\  
 38+ [Zn] &   $4s4d$&  $4s_{+},4d_{+})_{2} $&  14 &  2 453 320 &	 &   370  \\  
\\
 37+ [Ga] &   $4p$&  $4p_{-}$ &  1 &  0	& &  0   \\  
 37+ [Ga] &   $4p$&  $4p_{+}$ &  2 & 502 760	 &&  140$\dagger$   \\  
 37+ [Ga] &   $4s4p^{2}$&  $4s_{+}$ &  3 &  659 610	 &&  50   \\  
 37+ [Ga] &   $4s4p^{2}$&  $((4s_{+},4p_{-})_{0},4p_{+})_{3/2}$ &  4 &  	1 074 090 &&   140  \\  
 37+ [Ga] &   $4s4p^{2}$&  $((4s_{+},4p_{-})_{1},4p_{+})_{5/2}$ & 5 &  1 131 460	& &   240  \\  
 37+ [Ga] &   $4s4p^{2}$&  $((4s_{+},4p_{-})_{1},4p_{+})_{3/2}$ &  6 & 1 216 140	& &  170   \\  
 37+ [Ga] &   $4s4p^{2}$&  $((4s_{+},4p_{-})_{1},4p_{+})_{1/2}$ & 7 &  1 237 970	& &  220    \\  
 37+ [Ga] &   $4d$&  $4d_{-}$ & 9 &  1 619 300	 &&   320  \\  
\\
 36+ [Ge] &   $4p^{2}$&  $(4p_{-}^{2})_{0}$ & 1 &  0	 &&  0   \\  
 36+ [Ge] &   $4p^{2}$&  $(4p_{-},4p_{+})_{2}$ & 3 &  488 190& &  190   \\  
 36+ [Ge] &   $4s4p^{3}$&  $(4s_{+},4p_{+)_{1}}$ &  7&  1 216 340 & &  180   \\  
 36+ [Ge] &   $4p4d$&  $(4p_{-},4d_{-})_{1}$ &  11 &  1 638 250	& &  330   \\  
\\
 35+ [As] &   $4p^{3}$&  $(4p_{+}^{3})_{3/2}$ & 1 &  0	& &  0   \\  
 35+ [As] &   $4p^{3}$&  $(4p_{-},(4p_{+}^{2})_{2})_{3/2}$ &  2 &  440 827&+x &   FAC  \\  
 35+ [As] &   $4p^{3}$&  $(4p_{-},(4p_{+}^{2})_{2})_{5/2}$ & 3 &   	479 420&+x &   70  \\  
 35+ [As] &   $4p^{2}4d$&  $4d_{-}$ &  6 &  1 061 840&+x	 &  90   \\  
 35+ [As] &   $4s4p^{4}$&  $(4s_{+},(4p_{+}^2)_{2})_{5/2}$ &  7&  1 089 300&+x  &  50   \\  
 35+ [As] &   $4p^{2}4d$&  $((4p_{-},4p_{+})_{1},4d_{+})_{7/2}$ & 17 &   1 610 740&+x	 &  170   \\  
 35+ [As] &   $4p^{2}4d$&  $((4p_{-},4p_{+})_{1},4d_{-})_{5/2}$ &  20 &  	1 646 720 &&  310   \\  
 35+ [As] &   $4p^{2}4d$&  $((4p_{-},4p_{+})_{2},4d_{-})_{3/2}$ & 21 &  	1 652 980 &&  370   \\  
\\
 34+ [Se] &   $4p^{4}$&  $(4p_{+}^{4})_{2}$ &  1 &  0	 &&  0   \\  
 34+ [Se] &   $4p^{4}$&  $(4p_{+}^{2})_{0}$ & 2 &  63 215&+x	 &  FAC   \\  
 34+ [Se] &   $4p^{3}4d$&  $(4p_{+},4d_{-})_{3}$ & 9&  1 089 480&	 &   180  \\  
 34+ [Se] &   $4p^{3}4d$&  $(4p_{+},4d_{+})_{2}$ &  13&  1 249 340	& &  180   \\  
 34+ [Se] &   $4p^{3}4d$&  $(4p_{+},4d_{+})_{3}$ & 14 &  1 266 900	& &  190   \\  
 34+ [Se] &   $4p^{3}4d$&  $(4p_{+},4d_{+})_{1}$ & 15 &  1 320 360&+x	 &  190   \\  
 34+ [Se] &   $4p^{3}4d$&  $((4p_{-},(4p_{+}^{2})_{2})_{5/2},4d_{-})_{2}$ & 29&   1 660 300	& &   590  \\  
 34+ [Se] &   $4p^{3}4d$&  $((4p_{-},(4p_{+}^{2})_{2})_{3/2})_{3}$ &  30 &  1 668 630&	 &  490   \\  
\\
 33+ [Br] &   $4p^{5}$&  $(4p_{+}^{3})_{3/2}$ &  1 &  0	& &  0   \\  
 33+ [Br] &   $4p^{4}4d$&  $((4p_{+}^2)_{2},4d_{-})_{3/2}$ &  3 & 1 043 590	& &   130  \\  
 33+ [Br] &   $4p^{4}4d$&  $((4p_{+}^{2})_{2},4d_{-})_{5/2}$ & 5 &  1 056 800	 &&  530   \\  
 33+ [Br] &   $4p^{4}4d$&  $((4p_{+}^{2})_{2},4d_{-})_{7/2}$ & 6 & 1 077 240	& &   160  \\  
 33+ [Br] &   $4p^{4}4d$&  $((4p_{+}^{2})_{2},4d_{+})_{3/2}$ &  12& 1 270 380	 &&  210   \\  
 33+ [Br] &   $4s$	     &  $4s_{+}$ & 13 & 1 272 610	 & &  160   \\  
 33+ [Br] &   $4p^{4}4d$&  $((4p_{+}^{2})_{2},4d_{+})_{5/2}$ &  14&  1 288 160	 & &   190  \\  
 33+ [Br] &   $4p^{4}4d$&  $((4p_{-},(4p_{+}^{3})_{3/2})_{2},4d_{-})_{1/2}$ &  24&  1 672 540	 & &  190    \\  
 33+ [Br] &   $4p^{4}4d$&  $((4p_{-},(4p_{+}^{3})_{3/2})_{2},4d_{-})_{3/2}$ &  25&  1 672 540	 & &   190  \\  
 33+ [Br] &   $4p^{4}4d$&  $((4p_{-},(4p_{+}^{3})_{3/2})_{2},4d_{-})_{5/2}$ &  26&  1 681 280	 & &  170    \\  
\\
 32+ [Kr] &   $4p^{6}$&  $(4p_{+}^{4})_{0}$ &  1 &  0	 &&  0   \\  
 32+ [Kr] &   $4p^{5}4d$&  $((4p_{+}^{3})_{3/2},4d_{-})_{1}$ & 3 &  1 062 370	 &&   130  \\  
 32+ [Kr] &   $4p^{5}4d$&  $((4p_{+}^{3})_{3/2},4d_{+})_{1}$ &  9&  1 303 010	& &  200   \\  
 32+ [Kr] &   $4p^{5}4d$&  $(4p_{-},4d_{-})_{1}$ &  13&   1 684 740 & &   380  \\  
\\
 31+ [Rb] &   $4d$&  $4d_{-}$ & 1&  0	& &  0   \\  
 31+ [Rb] &   $4d$&  $4d_{+}$ & 2 &  95 819&+x	 &  FAC   \\  
 31+ [Rb] &   $4p^{5}4d^{2}$&  $((4p_{+}^{3})_{3/2}, (4d_{-}^{2})_{2})_{5/2}$ &  5 & 1 067 570	& &  150   \\  
 31+ [Rb] &   $4p^{5}4d^{2}$&  $((4p_{+}^{3})_{3/2}, (4d_{-}^{2})_{2})_{7/2}$ & 6 &  1 082 580&+x	 &  120    \\  
 31+ [Rb] &   $4p^{5}4d^{2}$&  $((4p_{+}^{3})_{3/2}, (4d_{-}^{0})_{2})_{3/2}$  & 8 &   1 124 800 &	 &   170  \\  
 31+ [Rb] &   $4p^{5}4d^{2}$&  $(((4p_{+}^{3})_{3/2},4d_{-})_{1},4d_{+})_{3/2}$ & 11 &  1 158 890&	 &  160  \\  
 31+ [Rb] &   $4p^{5}4d^{2}$&  $(((4p_{+}^{3})_{3/2},4d_{-})_{3},4d_{+})_{7/2}$ &  14 &  1 179 840&+x 	 &  140   \\  
 31+ [Rb] &   $4p^{5}4d^{2}$&  $(((4p_{+}^{3})_{3/2},4d_{-})_{2},4d_{+})_{7/2}$ &  16 &  	1 207 390 &+x &   150  \\  
 31+ [Rb] &   $4p^{5}4d^{2}$&  $(((4p_{+}^{3})_{3/2},4d_{-})_{3},4d_{+})_{5/2}$ &  17 & 1 211 910 	& &  180   \\  
 31+ [Rb] &   $4p^{5}4d^{2}$& $((4p_{+}^{3})_{3/2},(4d_{+}^{2})_{4})_{5/2}$  &  21 &   1 255 770	& &  200   \\  
 31+ [Rb] &   $4p^{5}4d^{2}$& $((4p_{+}^{3})_{3/2},(4d_{+}^{2})_{4})_{5/2}$  &  22 &  1 255 770	& &  200   \\  
 31+ [Rb] &   $4p^{5}4d^{2}$&  $(((4p_{+}^{3})_{3/2},4d_{-})_{3},4d_{+})_{3/2}$ & 25 &   1 279 690 &	 &   190  \\  
 31+ [Rb] &   $4p^{5}4d^{2}$&  $(((4p_{+}^{3})_{3/2},4d_{-})_{3},4d_{+})_{1/2}$ &  28 &  1 341 500&	 &  210  \\  
 31+ [Rb] &   $4f$&  $4f_{-}$ & 32 &   1 446 140	 &&  240   \\  
 31+ [Rb] &   $4p^{5}4d^{2}$&  $(4p_{-},(4d_{-}^2)_{2})_{5/2}$ & 34 &   1 589 690 &	 &   300  \\  
 31+ [Rb] &   $4p^{5}4d^{2}$&  $(4p_{-},(4d_{-}^{2})_{0})_{1/2}$ &  40 &  	 1 674 780 &&  510   \\  
 31+ [Rb] &   $4p^{5}4d^{2}$&  $(4p_{-},(4d_{+}^{2})_{2})_{5/2}$& 45 &  1 720 010	& &  400   \\  

\end{longtable}
\end{landscape}

\end{document}